\documentclass[12pt,english, leqno]{article}
\usepackage[T1]{fontenc}
\usepackage[latin1]{inputenc}
\usepackage{babel}
\usepackage{array}
\usepackage{verbatim}
\usepackage{multirow}
\usepackage{pdfpages}
\usepackage{pdflscape}
\usepackage{rotating}
\usepackage{tabularx}
\usepackage{standalone}
\usepackage{siunitx}
\usepackage{amsmath, amsthm, amsfonts, amssymb}
\usepackage{accents}
\usepackage{comment}
\usepackage{ragged2e}
\usepackage{colortbl}
\usepackage[flushleft]{threeparttable}
\usepackage[normalem]{ulem}
\usepackage[toc,page,header]{appendix}
\usepackage{placeins}
\usepackage{minitoc}
\usepackage{titletoc}
\usepackage{graphicx}
\usepackage[abs]{overpic}
\usepackage{setspace}
\usepackage[authoryear]{natbib}
\usepackage{multicol}
\usepackage{amsbsy}
\usepackage{caption}
\usepackage{lscape}
\usepackage[norule,splitrule]{footmisc}
\usepackage[capposition=top]{floatrow}
\usepackage{subcaption}
\usepackage{varwidth}
\usepackage{geometry}
\usepackage{dcolumn}
\usepackage{booktabs}
\usepackage{mathptmx}
\usepackage{nicefrac}
\usepackage{footnote}

\DeclareSymbolFont{newfont}{OML}{cmm}{m}{it}
\DeclareMathSymbol{\Varrho}{3}{newfont}{37}

\theoremstyle{definition}

\makeatletter

\newcolumntype{d}[1]{D{.}{.}{#1}}
\DeclareCaptionFormat{linebreak}{        \begin{varwidth}{\linewidth}    \centering
    #1#2#3  \end{varwidth}}
\makeatother

\begin{document}

\defcitealias{lmmp22}{LMMP (2022)} \defcitealias{StockYogo05}{SY (2005)}

\doparttoc
\faketableofcontents

\renewcommand{\thesection}{\Roman{section}} \renewcommand{\thesubsection}{%
\Roman{section}.\Alph{subsection}}

\pagestyle{plain}

\thispagestyle{empty}

\setcounter{footnote}{0}

\begin{titlepage}
\newgeometry{margin=1.2in}

\noindent \begin{center}
~\\
~\\
{\LARGE{}Robust Conditional Wald Inference for Over-Identified IV}\let\thefootnote\relax\footnotetext{* Lee: Princeton University and NBER (email: davidlee@princeton.edu); McCrary: Columbia University and NBER (email: jmccrary@law.columbia.edu); Moreira: FGV EPGE (email: mjmoreira@fgv.br); Porter: University of Wisconsin (email: jrporter@ssc.wisc.edu); Yap: Princeton University (email: lyap@princeton.edu).
}\\
\par\end{center}

\begin{center}
\ \\
\par\end{center}

\singlespacing

\begin{center}
{\large By David S. Lee, Justin McCrary, Marcelo J. Moreira, \\
Jack Porter, Luther Yap*}
\end{center}

\begin{center}
{\large{}November 2023}{\large\par}
\par\end{center}
\begin{abstract}
\begin{singlespace}

For the over-identified linear instrumental variables model, researchers commonly report the 2SLS estimate along with the robust standard error and seek to conduct inference with these quantities. If errors are homoskedastic, one can control the degree of inferential distortion using the first-stage $F$ critical values from \citet{StockYogo05}, or use the robust-to-weak instruments Conditional Wald critical values of \citet{Moreira03}. If errors are non-homoskedastic, these methods do not apply. We derive the generalization of Conditional Wald critical values that is robust to non-homoskedastic errors (e.g., heteroskedasticity or clustered
variance structures), which can also be applied to nonlinear weakly-identified models (e.g. weakly-identified GMM).

\bigskip{}

Keywords: Instrumental Variables, Weak Instruments, $t$-ratio, First-stage $F$-statistic, Conditional Wald

\end{singlespace}
\end{abstract}

\restoregeometry
\end{titlepage}

\setcounter{footnote}{0}

\onehalfspacing



\section{Introduction}

This paper considers inference in the over-identified linear instrumental
variables model and its generalization to weakly-identified models and GMM
more generally. The core problem of inference in the weak IV literature is
that when instruments are weak, conventional asymptotic approximations are
poor, causing standard inference procedures (like Wald or $t$-ratio-based
inference) to over-reject, even under the null. Indeed, \citet{Dufour97}
pointed out that any confidence set that is bounded with probability 1 (like
the usual $\hat{\beta}\pm1.96\cdot\hat{se}(\hat{\beta})$) could have the
potential to cover the true parameter 0 percent of the time (i.e., zero
percent confidence level).

\citet{Moreira03} provided a generalized approach to constructing inference
procedures that addressed this weak instrument problem via data-dependent
critical values; this approach was first demonstrated in the case when
errors are homoskedastic. \citet{Moreira03} presented, for the hypothesis
that the parameter is equal to a particular value, conditional versions of
the "trinity" of test procedures: Likelihood Ratio (LR), Lagrange Multiplier
(LM), and Wald tests. The critical values for each of these test statistics
were functions of the observed data and the null hypothesis.

Since then, a number of efforts have generalized these tests to accommodate
non-homoskedastic settings, given the widespread preference of applied
researchers to remain somewhat agnostic about the properties of the errors
in the linear model.\footnote{%
See, for example, \citet{AndrewsMoreiraStock04}, \citet{Kleibergen05}, %
\citet{Andrews16}.} Curiously, while there have been efforts to generalize
the Conditional LR (CLR) and LM tests to general non-homoskedastic errors,
to the best of our knowledge, the extension of the Conditional Wald in such
a way has been neglected.

In this paper, we derive the extension of Conditional Wald to
non-homoskedastic settings. This effort delivers a Wald-based inference
procedure that is valid, similar, robust to arbitrarily weak instruments,
robust to HAC error structures, and applicable to more general
weakly-identified settings like GMM.

There are a number of practical reasons to revisit a testing procedure
rooted in a Wald approach. First, for the linear instrumental variables
model, applied research has revealed a preference for Wald-based inference.
Most typically, researchers compute and report the 2SLS estimator and robust
standard errors, regardless of concerns about instrument weakness. In
homoskedastic settings, researchers could pursue two different options for
conducting Wald-based inference using the computed $t$-ratio. Researchers
can either use the critical value function for Conditional Wald in %
\citet{Moreira03}, or they can use the first-stage $F$-statistic along with
the critical value tables in \citet{StockYogo05} and the Bonferroni
arguments used in \hbox{\citet{StaigerStock97}}. When the errors are
non-homoskedastic -- as is typically allowed in modern empirical work -- the
values in \citet{StockYogo05} tables no longer reliably control size
distortions, as pointed out in \citet{AndrewsStockSun19}.  The contribution of the current paper is to
provide a method for computing critical values for the $t$-ratio that will
deliver valid, robust inference, even in non-homoskedastic settings.\footnote{In this paper, acceptance/rejection of the null is the result of comparing a single statistic with a valid (in this case, data-dependent) critical value. A different, "two-step" inference approach where two different procedures (one "robust to weak instruments" and the other non-robust) are combined to form an overall valid procedure (which can also accommodate non-homoskedastic settings) is proposed by \citet{Andrews18}.
}

A second reason to consider a robust-to-HAC version of Conditional Wald of %
\citet{Moreira03} is that there are two recent studies pointing to power
advantages of Conditional Wald in the homoskedastic, over-identified setting
and in the non-homoskedastic just-identified setting. 
\citet{vandesijpewindmeijer2023} analyze power for the over-identified,
homoskedastic case, and provide simulation evidence that Conditional Wald
using 2SLS tends to produce shorter confidence set lengths, compared to CLR (%
\citet{Moreira03}). This is a particularly striking finding, in light of
papers that point to the near-optimality, in terms of power, of CLR.
Furthermore, \citet{lmmpy23}, in the context of the just-identified (robust
to HAC errors) IV model, show that two different Wald-based -- $VtF$ and
Conditional Wald -- confidence intervals appear to be almost always shorter
than that of \citet{AndersonRubin49}, a recommended benchmark in the
literature. Thus, developing the Conditional Wald robust to HAC errors is
not only already aligned with practitioner practice, but these recent
studies suggest that it may even have power advantages in the form of
shorter confidence intervals.

Our motivation for deriving CW critical values is entirely practical and
stems from taking as given practitioners' apparent preference for computing
the 2SLS point estimate and robust standard error (presuming
non-homoskedasticity), and finding critical values that lead to valid
inference. Our approach is thus different from identifying the optimal test
after having defined a class of procedures and an objective function.
Nevertheless, the two studies mentioned above do suggest the possibility
that in terms of power and confidence interval length, CW could fare well
compared to existing alternatives for the over-identified model.

Section \ref{ivmodel} establishes the notation we use for the standard linear IV model with non-homoskedastic errors, Section \ref{tests} derives the critical values for Robust Conditional Wald tests based on 2SLS, LIML, two-step, and CUE GMM estimators, Section \ref{sec:generalization} extends the test to nonlinear weakly-identified models (e.g. GMM), and Section \ref{conclusion} concludes.

\section{The Linear IV Model}\label{ivmodel}

\label{sec:linearIV}

The standard linear IV model is represented by
\begin{eqnarray*}
y_{1} &=&Y_{2}\beta +u \\
Y_{2} &=&Z\Pi +V_{2}
\end{eqnarray*}%
where $y_{1}$ $\left( n\times 1\right) $ is the dependent variable, $Y_{2}$ $%
\left( n\times p\right) $ are the endogenous variable(s) of interest, and $Z$
$\left( n\times k\right) $ are the excluded instruments, while $u$ $\left(
n\times 1\right) $ and $V_{2}$ $\left( n\times p\right) $ are the unobserved
structural-form errors. The single endogenous regressor case simply
corresponds to $p=1$. We will always take $k\geq p$ with $k=p$ corresponding
to the just-identified model and $k>p$ corresponding to the over-identified
model. The parameter of interest is $\beta $. It is straightforward to
accommodate additional covariates (including a constant), but we omit their
inclusion in the exposition below.

The reduced-form model is:
\begin{eqnarray*}
y_{1} &=&Z\Pi \beta +v_{1} \\
Y_{2} &=&Z\Pi +V_{2},
\end{eqnarray*}%
where $u\equiv v_{1}-V_{2}\beta $. It will be convenient to write the model
in a matrix form:%
\begin{equation*}
Y=Z\Pi A+V,
\end{equation*}%
where $Y=\left[ y_{1}:Y_{2}\right] $, $V=\left[ v_{1}:V_{2}\right] $, and $A=%
\left[ \beta :I_{p}\right] $. We will use the notation $Y_{i}$, $V_{i}$, $%
Z_{i}$, etc, to denote the $i$-th row of the corresponding matrix.

In \citet{Moreira03}, the rows of $V$ were assumed to be i.i.d. This paper
relaxes this assumption for the derivation of the Conditional Wald test
robust to different DGPs. We are motivated by the observation that applied
researchers typically prefer not to make the assumption of homoskedasticity,
and often they are interested in a clustered error structure, for example.

\section{The Robust Conditional Wald Tests}\label{tests}

In this section, we derive the Conditional Wald (CW) tests robust to HAC
errors for the linear model given in section~\ref{sec:linearIV}.

The Wald statistic is formed by three elements: a null value $\beta _{0}$,
an estimator $\widehat{\beta }_{n}$, and a robust asymptotic variance
estimator $\widehat{A.Var}$ for $\sqrt{n}\left( \widehat{\beta }_{n}-\beta
_{0}\right) $:
\begin{equation*}
\widehat{\mathcal{W}}_{n}=n\left( \widehat{\beta }_{n}-\beta _{0}\right)
^{\prime }\left[ \widehat{A.Var}\right] ^{-1}\left( \widehat{\beta }%
_{n}-\beta _{0}\right) .
\end{equation*}%
In section~\ref{subsec:estimators}, we review the class of linear GMM
estimators for $\beta $, which includes common estimators like 2SLS, LIML,
efficient two-step GMM, and the CU (continuously updating) GMM estimator, all of which can be used for constructing a Robust Conditional Wald test. In
section~\ref{subsec:test}, we review the robust variance estimators based on the
asymptotic distribution of $\sqrt{n}\left( \widehat{\beta }_{n}-\beta
_{0}\right) $. With these components in hand, we can form robust versions of
the Wald statistic for various estimators, $\widehat{\beta }_{n}$. Note that there are no new results in Sections \ref{subsec:estimators} and \ref{subsec:test}, and there are many references that detail these standard results (as one example, see \citet{NeweyMcFadden94}). We review a selected set of well-established facts about GMM to highlight that the Robust Conditional Wald test we derive in \ref{subsec:cvf} is not specific to the leading case in applied work -- 2SLS -- and can be applied to tests based on other estimators of the parameter of interest. Note that the multitude of different estimators that could be employed arises in the over-identified case; in contrast, for example, in the single instrument case, all of the estimators we discuss below collapse to the standard IV estimator. 

With that as context, an interested reader can skip to  section~\ref{subsec:cvf}, in which we show how to apply the conditional argument of \citet{Moreira03} to obtain a critical value function for the Wald
statistic that is robust to instrument weakness. The critical value function
is then used to form a robust similar test, which can be inverted to
generate confidence intervals for $\beta$.

\subsection{Estimators}

\label{subsec:estimators}

Estimators like 2SLS or LIML can be viewed as particular GMM estimators based on the linear moment
condition:%
\begin{equation*}
\overline{g}_{n}\left( \beta \right) =n^{-1}\sum\limits_{i=1}^{n}Z_{i}\left(
y_{1i}-Y_{2i}^{\prime }\beta \right) =n^{-1}Z^{\prime }Yb,
\end{equation*}%
where $b=\left( 1,-\beta ^{\prime }\right) ^{\prime }$. A GMM estimator for $%
\beta $ is the minimizer of the criterion%
\begin{equation}
\widehat{Q}_{n}\left( \beta \right) =\overline{g}_{n}\left( \beta \right)
^{\prime }W_{n}\left( \beta \right) \overline{g}_{n}\left( \beta \right)
\label{(criterion)}
\end{equation}%
where the weighting matrix may or may not depend on the unknown coefficient $%
\beta $. Different choices of $W_{n}\left( \beta \right) $ will lead to different estimators $\widehat{\beta }_{n}$.

For the 2SLS estimator, the weighting matrix does not depend on $%
\beta $:
\begin{equation*}
W_{n}{}^{-1}=\widehat{V}_{u}\cdot n^{-1}\sum\limits_{i=1}^{n}Z_{i}Z_{i}^\prime=%
\widehat{V}_{u}\cdot n^{-1}Z^{\prime }Z,
\end{equation*}%
where $\widehat{V}_{u}$ is an estimator of $V_{u}$, which is the variance of $u$. Because $%
\widehat{Q}_{n}\left( \beta \right) $ is quadratic in $\beta $, it is
straightforward to show that the resulting estimator is 2SLS:
\begin{equation}
\widehat{\beta }_{n}=\left[ Y_{2}^{\prime }Z\left( Z^{\prime }Z\right)
^{-1}Z^{\prime }Y_{2}\right] ^{-1}Y_{2}^{\prime }Z\left( Z^{\prime }Z\right)
^{-1}Z^{\prime }y_1.  \label{(2SLS estimator)}
\end{equation}%
We can re-express the estimator as%
\begin{equation*}
\widehat{\beta }_{n}=\left[ \widehat{Y}_{2}^{\prime }\widehat{Y}_{2}\right]
^{-1}\widehat{Y}_{2}^{\prime }y_1,
\end{equation*}%
where $\widehat{Y}_{2}=NY_{2}$ and $N=Z\left( Z^{\prime }Z\right)
^{-1}Z^{\prime }$ is the usual projection matrix. That is, in the first
stage, we first regress $Y_{2}$ on $Z$ to obtain the fitted values $\widehat{%
Y}_{2}$. In the second stage, we regress $y_{1}$ on the fitted values $%
\widehat{Y}_{2}$.

For the LIML estimator, the weight matrix is formed by using $\beta $ along with
residuals from OLS regressions of $y_{1}$ and $Y_{2}$ on $Z$. Let $b=\left( 1,-\beta ^{\prime }\right) ^{\prime }$, $%
N=Z\left( Z^{\prime }Z\right) ^{-1}Z^{\prime }$, and $M=I-N$. Then%
\begin{equation*}
\widehat{V}_{u}\left( \beta \right) =n^{-1}\sum\limits_{i=1}^{n}\left(
\widehat{v}_{1,i}-\widehat{V}_{2,i}^{\prime }\beta \right)
^{2}=n^{-1}b^{\prime }Y^{\prime }MYb,
\end{equation*}%
where $\widehat{V}=MY=MV$. The GMM\ criterion is no longer quadratic in $%
\beta $ once we use
\begin{equation*}
W_{n}\left( \beta \right) ^{-1}=\widehat{V}_{u}\left( \beta \right) \cdot
n^{-1}\sum\limits_{i=1}^{n}Z_{i}Z_{i}^\prime=\widehat{V}_{u}\left( \beta \right)
\cdot n^{-1}Z^{\prime }Z.
\end{equation*}%
Instead it is a ratio of quadratic forms:%
\begin{equation*}
\widehat{Q}_{n}\left( \beta \right) =\frac{n^{-1}\sum\limits_{i=1}^{n}\left(
y_{1i}-Y_{2i}^{\prime }\beta \right) Z_{i}\left(
n^{-1}\sum\limits_{i=1}^{n}Z_{i}Z_{i}^{\prime }\right)
^{-1}n^{-1}\sum\limits_{i=1}^{n}Z_{i}\left( y_{1i}-Y_{2i}^{\prime }\beta
\right) }{n^{-1}\sum\limits_{i=1}^{n}\left( \widehat{V}_{1,i}-\widehat{V}%
_{2,i}^{\prime }\beta \right) ^{2}}.
\end{equation*}%
The minimum of
\begin{equation*}
\widehat{Q}_{n}\left( \beta \right) =\frac{b^{\prime }Y^{\prime }NYb}{%
b^{\prime }Y^{\prime }MYb}
\end{equation*}%
is well-known to lead to the LIML\ estimator (see %
\citet{DavidsonMacKinnon21} among others). This estimator is proportional to
the eigenvector associated to the smallest eigenvalue $\underline{\lambda }%
_{n}$of the characteristic polynomial $\left\vert Y^{\prime }NY-\lambda
.Y^{\prime }MY\right\vert =0$.

Once we consider different weighting functions $W_{n}\left( \beta \right) $,
we can wonder if there is the \textquotedblleft best\textquotedblright\
possible choice. The answer depends if errors are heteroskedastic,
clustered, etc. Only in special cases, such as with homoskedastic errors,
are the 2SLS and LIML\ estimators \textquotedblleft best.\textquotedblright\
To obtain the weighting function that optimally accounts for  heteroskedasticity,
clustering, serial correlation, and other departures from homoskedastic
errors with no serial correlation, one first considers the (infeasibly estimated) variance of
\begin{equation*}
vec\left( n^{-1/2}\sum\limits_{i=1}^{n}Z_{i}V_{i}^{\prime }\right)
=n^{-1/2}\sum\limits_{i=1}^{n}\left( V_{i}\otimes Z_{i}\right) .\footnote{%
Because the $Z_{i}V_{i}^{\prime }$ is an $k\times p$ matrix, we can stack
its columns to form a single vector with the $vec\left( \cdot \right) $
operator.}
\end{equation*}%
Under general conditions for the DGPs, the limiting variance exists and is
given by%
\begin{equation*}
\Omega =\lim_{n} \, n^{-1}\sum\limits_{i=1}^{n}\sum\limits_{j=1}^{n}C\left(
V_{i}\otimes Z_{i},V_{j}\otimes Z_{j}\right).
\end{equation*}

Since we do not observe the errors $V_{i}$, we can make this feasible by replacing them with, as an example, the OLS\ residuals $%
\widehat{V}_{i}$. There are different estimators for the variances and
covariances above, each one of them suited to different assumptions on the
DGPs. For example, typically, one uses the variance estimate of \citet{White80} for
heteroskedastic errors that are serially uncorrelated:%
\begin{equation*}
\widehat{\Omega }_{n}=n^{-1}\sum\limits_{i=1}^{n}\left( \widehat{V}%
_{i}\otimes Z_{i}\right) \left( \widehat{V}_{i}\otimes Z_{i}\right) ^{\prime
}.
\end{equation*}%
Henceforth, we will employ the broader notation $\widehat{\Omega }_{n}$
without explicitly specifying its formulae for different departures from
homoskedasticity. Examples of robust variance estimators include %
\citet{White80} for heteroskedasticity, \citet{NeweyWest87} and %
\citet{Andrews91} for both heteroskedasticity and autocorrelation (HAC), and %
\citet{CameronGelbachMiller11} for clustered errors. See %
\citet{AndrewsMoreiraStock04} for the IV model.

The GMM criterion is then
\begin{eqnarray*}
\widehat{Q}_{n}\left( \beta \right) &=& \left[ n^{-1}\sum\limits_{i=1}^{n}%
\left( Y_{i}-X_{i}^{\prime }\beta \right) Z_{i} \right] W_{n} \left[
n^{-1}\sum\limits_{i=1}^{n}Z_{i}\left( Y_{i}-X_{i}^{\prime }\beta \right) %
\right] \text{, where} \\
W_{n}{}^{-1} &=&\left( \widetilde{b}_{n}\otimes I_{k}\right) ^{\prime }%
\widehat{\Omega }_{n}\left( \widetilde{b}_{n}\otimes I_{k}\right) \text{ and
}\widetilde{b}_{n}=\left( 1,-\widetilde{\beta }_{n}^{\prime }\right)
^{\prime }\text{,}
\end{eqnarray*}%
with $\widetilde{\beta }_{n}$ being a preliminary consistent estimator of $%
\beta $. Again, this criterion
\begin{equation*}
\widehat{Q}_{n}\left( \beta \right) =b^{\prime }Y^{\prime }ZW_{n}Z^{\prime
}Yb
\end{equation*}%
is quadratic in $\beta $ and we can easily find its closed-form solution:%
\begin{equation}
\widehat{\beta }_{n}=\left[ Y_{2}^{\prime }ZW_{n}Z^{\prime }Y_{2}\right]
^{-1}Y_{2}{}^{\prime }ZW_{n}Z^{\prime }y.  \label{(two-step GMM estimator)}
\end{equation}%
which is the two-step GMM\ estimator. It simplifies to the
2SLS estimator if $W_{n}^{-1}$ is proportional to $n^{-1}Z^{\prime }Z$. When the weight matrix depends on $\beta $, we
obtain%
\begin{eqnarray*}
\widehat{Q}_{n}\left( \beta \right) &=&n^{-1}\sum\limits_{i=1}^{n}\left(
y_{1i}-Y_{2i}^{\prime }\beta \right) Z_{i}W_{n}{}\left( \beta \right)
n^{-1}\sum\limits_{i=1}^{n}Z_{i}\left( y_{1i}-Y_{2i}^{\prime }\beta \right)
\text{, where} \\
W_{n}{}\left( \beta \right) ^{-1} &=&\left( b\otimes I_{k}\right) ^{\prime }%
\widehat{\Omega }_{n}\left( b\otimes I_{k}\right) .
\end{eqnarray*}%
which is the Continuously Updating (CU) GMM estimator,
proposed by \citet{HansenSingleton82}.


\subsection{Wald Test Statistics}

\label{subsec:test}

Finally, the usual Wald test statistics are based on the standard asymptotic approximation to the distribution of the GMM estimators. Under the true parameter $\beta _{0}$,%
\begin{equation*}
n^{1/2}\overline{g}_{n}\left( \beta _{0}\right) \rightarrow _{d}N\left(
0,V_{0}\right) \text{, where }V_{0}=\left( b_{0}\otimes I_{k}\right)
^{\prime }\Omega \left( b_{0}\otimes I_{k}\right)
\end{equation*}%
for $b_{0}=\left( 1,-\beta _{0}^{\prime }\right) $. For convenience, we
derive the asymptotic distribution where we use the parameter $\beta _{0}$
in the criterion function:%
\begin{equation*}
\overline{Q}_{n}\left( \beta \right) =\overline{g}_{n}\left( \beta \right)
^{\prime }W_{n}\left( \beta _{0}\right) \overline{g}_{n}\left( \beta \right)
.
\end{equation*}%
%
%
%
%
%
%
%
%
%
%
%
%
This setup allows for the possibility that the limiting behavior of $%
W_{n}\left( \beta _{0}\right) $ is not necessarily proportional to $%
V_{0}^{-1}$. Hence, $\widehat{\beta }_{n}$ is not necessarily optimal. Under
the usual asymptotics, the distribution of estimators which minimize $%
\overline{Q}_{n}\left( \beta \right) $ is the same as if we had used $%
\widehat{Q}_{n}\left( \beta \right) $ instead, where the weight uses a
preliminary estimator or uses $\beta $ itself (derivations of these results are standard and can be found, e.g. in \citet{NeweyMcFadden94}). As a result, the 2SLS and LIML
estimators are asymptotically equivalent, while the two-step GMM\ and CUE
estimators are asymptotically equivalent as well. We derive the asymptotic
distribution for the 2SLS and two-step GMM estimators from equations (\ref%
{(2SLS estimator)}) and (\ref{(two-step GMM estimator)}) and, so, for the
LIML\ and continuously updating estimators as well.

For the 2SLS estimator, we can write%
\begin{equation*}
n^{1/2}\left( \widehat{\beta }_{n}-\beta _{0}\right) =\left[
n^{-1}Y_{2}^{\prime }Z\left( n^{-1}Z^{\prime }Z\right) ^{-1}n^{-1}Z^{\prime
}Y_{2}\right] ^{-1}n^{-1}Y_{2}^{\prime }Z\left( n^{-1}Z^{\prime }Z\right)
^{-1}n^{-1/2}Z^{\prime }Vb.
\end{equation*}%
Assuming that the following probability limits exist,%
\begin{equation*}
plim\text{ }n^{-1}Y_{2}^{\prime }Z=E_{Y_{2}^{\prime }Z}\text{ and }plim\text{
}n^{-1}Z^{\prime }Z=E_{Z^{\prime }Z}\text{,\footnote{%
If the process for $\left( y_{1,i},Y_{2,i},Z_{i}\right) $ were ergodic with
enough moments, then $E_{Z^{\prime }Z}=E\left( Z_{i}Z_{i}^{\prime }\right) $
and $E_{Y_{2}^{\prime }Z}=E\left( Y_{2i}Z_{i}^{\prime }\right) $. However,
this notation allows for deterministic values of IVs as well as clustering.}}
\end{equation*}%
we then have $n^{1/2}\left( \widehat{\beta }_{n}-\beta _{0}\right)
\rightarrow _{d}N\left( 0,B_{0}^{-1}A_{0}B_{0}^{-1}\right) $, where%
\begin{eqnarray*}
B_{0} &=&\left[ E_{Y_{2}^{\prime }Z}E_{Z^{\prime }Z}^{-1}E_{Z^{\prime }Y_{2}}%
\right] ^{-1}\text{ and} \\
A_{0} &=&E_{Y_{2}^{\prime }Z}E_{Z^{\prime }Z}^{-1}\left( b_{0}\otimes
I_{k}\right) ^{\prime }\Omega \left( b_{0}\otimes I_{k}\right) E_{Z^{\prime
}Z}^{-1}E_{Z^{\prime }Y_{2}}\text{.}
\end{eqnarray*}%
We can find some consistent estimators for $A_{0}$ and $B_{0}$, and derive a
Wald statistic for the 2SLS and LIML\ estimators:%
\begin{eqnarray*}
\widehat{\mathcal{W}}_{n}^{\ast } &=&n\left( \widehat{\beta }_{n}-\beta
_{0}\right) ^{\prime }\left[ \widehat{B}_{n}^{-1}\widehat{A}_{n}\widehat{B}%
_{n}^{-1}\right] ^{-1}\left( \widehat{\beta }_{n}-\beta _{0}\right) \text{,
where} \\
\widehat{B}_{n} &=&n^{-1}Y_{2}^{\prime }Z\left( Z^{\prime }Z\right)
^{-1}Z^{\prime }Y_{2}\text{ and} \\
\widehat{A}_{n} &=&Y_{2}^{\prime }Z\left( Z^{\prime }Z\right) ^{-1}\left(
\widehat{b}_{n}\otimes I_{k}\right) ^{\prime }\widehat{\Omega }_{n}\left(
\widehat{b}_{n}\otimes I_{k}\right) \left( Z^{\prime }Z\right)
^{-1}Z^{\prime }Y_{2}\text{,}
\end{eqnarray*}%
with $\widehat{b}_{n}=\left( 1,-\widehat{\beta }_{n}^{\prime }\right)
^{\prime }$ based on the respective 2SLS/LIML\ estimator.\footnote{%
We typically use $\left( \widehat{b}_{n}\otimes I_{k}\right) ^{\prime }%
\widehat{\Omega }_{n}\left( \widehat{b}_{n}\otimes I_{k}\right) $ as a
consistent estimator for $V_{0}$. However, other estimators are possible,
including $\left( b_{0}\otimes I_{k}\right) ^{\prime }\widehat{\Omega }%
_{n}\left( b_{0}\otimes I_{k}\right) $.}

Likewise, for the two-step GMM\ estimator, we find that $n^{1/2}\left(
\widehat{\beta }_{n}-\beta _{0}\right) \rightarrow _{d}N\left(
0,B_{0}^{-1}\right) $, where
\begin{equation*}
B_{0}=E_{Y_{2}^{\prime }Z}\left[ \left( b_{0}\otimes I_{k}\right) ^{\prime
}\Omega \left( b_{0}\otimes I_{k}\right) \right] ^{-1}E_{Z^{\prime }Y_{2}}.
\end{equation*}%
We can find a consistent estimator for $B_{0}$ and derive a Wald statistic
for the two-step GMM\ and continuously updating estimator:%
\begin{eqnarray*}
\widehat{\mathcal{W}}_{n}^{o} &=&n\left( \widehat{\beta }_{n}-\beta
_{0}\right) ^{\prime }\widehat{B}_{n}\left( \widehat{\beta }_{n}-\beta
_{0}\right) \text{, where} \\
\widehat{B}_{n} &=&n^{-1}Y_{2}^{\prime }Z\left[ \left( \widehat{b}%
_{n}\otimes I_{k}\right) ^{\prime }\widehat{\Omega }_{n}\left( \widehat{b}%
_{n}\otimes I_{k}\right) \right] ^{-1}n^{-1}Z^{\prime }Y_{2}.
\end{eqnarray*}%
with $\widehat{b}_{n}=\left( 1,-\widehat{\beta }_{n}^{\prime }\right)
^{\prime }$ based on the GMM/CU estimators.\footnote{%
We can also use here either the null value $\beta _{0}$ or the preliminary
estimator $\widetilde{\beta }_{n}$ for the variance estimator.}

\subsection{Valid Critical Value Functions}\label{subsec:cvf}

As emphasized in \citet{Dufour97}, since the nuisance parameter representing the strength of the first stage may be arbitrarily close to zero, then the usual constant critical values cannot be valid; indeed, \citet{Dufour97} points out that any valid confidence set in this context must be unbounded with positive probability, which clearly cannot be the case with a constant critical value for any of the Wald statistics mentioned above. To derive valid critical values, using the conditioning strategy of \citet{Moreira03}, we begin by defining the quantity%
\begin{equation*}
R=\left( Z^{\prime }Z\right) ^{-1/2}Z^{\prime }Y=\left[ R_{1}:R_{2}\right] ,
\end{equation*}%
where the $k$-dimensional vector $R_{1}$ is the first column of $R$ and the $%
k\times p$-matrix $R_{2}$ is the last $p$ columns of $R$. The
standardization avoids multiplication by the sample size $n$. The asymptotic
variance of $vec(R)$ is%
\begin{equation*}
\Sigma =\left( I_{p+1}\otimes \left( E_{Z^{\prime }Z}\right) ^{-1/2}\right)
\Omega \left( I_{p+1}\otimes \left( E_{Z^{\prime }Z}\right) ^{-1/2}\right) =
\left[
\begin{array}{cc}
\Sigma _{11} & \Sigma _{12} \\
\Sigma _{21} & \Sigma _{22}%
\end{array}%
\right] ,
\end{equation*}%
where the matrix $\Sigma $ is being partitioned by submatrices of
columns/rows of dimensions $1$ and $p$. Analogously, we can use the estimator%
\begin{equation*}
\widehat{\Sigma }_{n}=\left( I_{p+1}\otimes \left( n^{-1}Z^{\prime }Z\right)
^{-1/2}\right) \widehat{\Omega }_{n}\left( I_{p+1}\otimes \left(
n^{-1}Z^{\prime }Z\right) ^{-1/2}\right) =\left[
\begin{array}{cc}
\widehat{\Sigma }_{11,n} & \widehat{\Sigma }_{12,n} \\
\widehat{\Sigma }_{21,n} & \widehat{\Sigma }_{22,n}%
\end{array}%
\right] .
\end{equation*}

Up to a scale of the sample size $n$, the GMM criterion is%
\begin{equation*}
\widehat{Q}_{n}\left( \beta \right) =b^{\prime }R^{\prime }\left(
n^{-1}Z^{\prime }Z\right) ^{1/2}W_{n}\left( \beta \right) \left(
n^{-1}Z^{\prime }Z\right) ^{1/2}Rb=b^{\prime }R^{\prime }\overline{W}%
_{n}\left( \beta \right) Rb,
\end{equation*}%
where $\overline{W}_{n}\left( \beta \right) =\left( n^{-1}Z^{\prime
}Z\right) ^{1/2}W_{n}\left( \beta \right) \left( n^{-1}Z^{\prime }Z\right)
^{1/2}$. It is clear that only $R$ and the weight function $\overline{W}%
_{n}\left( \beta \right) $ fully determine the estimator $\widehat{\beta }%
_{n}$. To illustrate this connection, recall that the 2SLS estimator results
if $W_{n}^{-1}$ is proportional to $n^{-1}Z^{\prime }Z$. For such a weight,
we have $\overline{W}_{n}\left( \beta \right) =I_{k}$, and we trivially have
the 2SLS being dependent only on $R$. Indeed, the 2SLS estimator can be
written as%
\begin{equation*}
\widehat{\beta }=\left( R_{2}^{\prime }R_{2}\right) ^{-1}R_{2}^{\prime
}R_{1}.
\end{equation*}%
The same holds for the other estimators as well. For example, we take the
LIML\ estimator. When $W_{n}\left( \beta \right) ^{-1}=\widehat{V}_{u}\left(
\beta \right) \cdot n^{-1}Z^{\prime }Z$, we have $\overline{W}_{n}\left(
\beta \right) =\widehat{V}_{u}\left( \beta \right) ^{-1}I_{k}$. The LIML\
estimator solves%
\begin{equation*}
\widehat{Q}_{n}\left( \beta \right) =\frac{b^{\prime }R^{\prime }Rb}{%
b^{\prime }\widehat{\Phi }_{n}b}\text{, where }\widehat{\Phi }%
_{n}=n^{-1}Y^{\prime }MY.
\end{equation*}

Having found that the estimators are completely determined by the
standardized reduced-form coefficients $R$ and the function $\overline{W}%
_{n}\left( \beta \right) $, we can turn our attention to the Wald statistics.

The Wald statistic for the 2SLS/LIML\ estimators has the form%
\begin{equation*}
\widehat{\mathcal{W}}_{n}^{\ast }=\left( \widehat{\beta }_{n}-\beta
_{0}\right) ^{\prime }\left[ \left( R_{2}^{\prime }R_{2}\right)
^{-1}R_{2}^{\prime }\left( \widehat{b}_{n}\otimes I_{k}\right) ^{\prime }%
\widehat{\Sigma }_{n}\left( \widehat{b}_{n}\otimes I_{k}\right) R_{2}\left(
R_{2}^{\prime }R_{2}\right) ^{-1}\right] ^{-1}\left( \widehat{\beta }%
_{n}-\beta _{0}\right) .
\end{equation*}%
Hence, it is a function of $R$ and $\widehat{\Sigma }_{n}$ (or, for LIML, $%
\widehat{\Phi }_{n}$). Likewise, the Wald statistic for the two-step GMM\
and CU\ estimators can be written as
\begin{equation*}
\widehat{\mathcal{W}}_{n}^{o}=\left( \widehat{\beta }_{n}-\beta _{0}\right)
^{\prime }\left[ R_{2}^{\prime }\left[ \left( \widehat{b}_{n}\otimes
I_{k}\right) ^{\prime }\widehat{\Sigma }_{n}\left( \widehat{b}_{n}\otimes
I_{k}\right) \right] ^{-1}R_{2}\right] \left( \widehat{\beta }_{n}-\beta
_{0}\right) ,
\end{equation*}%
which again depends only on $R$ and $\widehat{\Sigma }_{n}$ (as long as the
preliminary estimator $\widetilde{\beta }_{n}$ depends only on $R$ and $%
\widehat{\Sigma }_{n}$ as well, such as the 2SLS estimator). In short, the Wald statistics associated with any of the estimators we have discussed above are functions of $R$, $\widehat{\Sigma }_{n}$,
and $\widehat{\Phi }_{n}$ as shown above.

We now apply the conditioning approach of \citet{Moreira03}, beginning by finding a useful transformation of $R$:%
\begin{equation*}
R_{0}=RB_{0}=\left[ R_{u}:R_{2}\right] \text{, where }B_{0}=\left[
\begin{array}{cc}
1 & 0_{1\times p} \\
-\beta _{0} & I_{p}%
\end{array}%
\right] .
\end{equation*}%
That is, $R_{u}=R_{1}-R_{2}\beta _{0}$. Note that the asymptotic variance of
$R_{0}$ is given by%
\begin{equation*}
\Sigma _{0}=\left( B_{0}^{\prime }\otimes I_{k}\right) \Sigma \left(
B_{0}\otimes I_{k}\right) =\left[
\begin{array}{cc}
\Sigma _{uu} & \Sigma _{u2} \\
\Sigma _{2u} & \Sigma _{22}%
\end{array}%
\right] .
\end{equation*}%
This quantity can of course be consistently estimated as well (regardless of
identification of $\beta $):%
\begin{equation*}
\widehat{\Sigma }_{0,n}=\left( B_{0}^{\prime }\otimes I_{k}\right) \widehat{%
\Sigma }_{n}\left( B_{0}\otimes I_{k}\right) =\left[
\begin{array}{cc}
\widehat{\Sigma }_{uu,n} & \widehat{\Sigma }_{u2,n} \\
\widehat{\Sigma }_{2u,n} & \widehat{\Sigma }_{22,n}%
\end{array}%
\right] .
\end{equation*}%
Consider a transformation of $R$.%
\begin{equation*}
\widehat{D}=vec\left( R_{2}\right) -\widehat{\Sigma }_{2u,n}\widehat{\Sigma }%
_{uu,n}^{-1}R_{u}.
\end{equation*}%
Given $\widehat{\Sigma }_{n}$, there is a one-to-one transformation between
the pair $R$ and $R_{2}$ and the pair $R_{u}$ and $\widehat{D}$. Since we have established that all of the Wald statistics above can be written as functions of $R, \widehat{\Sigma}_n,$ and $\widehat{\Phi}_n$, this means that they can also be written as functions of $R_u,  \widehat{D},  \widehat{\Sigma}_n, $ and $\widehat{\Phi}_n$. Importantly, adopting the appropriate assumptions relevant for HAC (e.g. see  \citet{Kleibergen05} or %
\citet{Andrews16}), it can be shown that $\left( R_{u},\widehat{D}\right) \rightarrow
_{d}\left( \mathcal{R}_{u},\mathcal{D}\right) $, where $\mathcal{R}_{u}$ and
$\mathcal{D}$ are asymptotically normal and independent, with $\mathcal{R}_{u}$ being mean zero with a variance matrix that can be consistently estimated under the null -- that is, 
$\mathcal{R}_{u}\sim N\left( 0,\Sigma _{uu}\right) $ under the null. As \citet{Moreira03} shows, this allows one to establish the distribution of test statistics even in the presence of the unknown nuisance parameter (the mean of $R_2$), since the distribution of $\mathcal{R}_{u}$ conditional on $\mathcal{D}$ is the same as the marginal distribution.\footnote{%
We will not standardize here the $R_{u}$ and $D$ statistics. However, we
could have worked with their respective standardized versions, $S=\left[
\left( b_{0}^{\prime }\otimes I_{k}\right) \Sigma \left( b_{0}\otimes
I_{k}\right) \right] ^{-1/2}Rb_{0}$ and $T=\left[ \left( A_{0}^{\prime
}\otimes I_{k}\right) \Sigma ^{-1}\left( A_{0}\otimes I_{k}\right) \right]
^{-1/2}\left( A_{0}^{\prime }\otimes I_{k}\right) \Sigma ^{-1}vec\left(
R\right) $, as in \citet{MoreiraMoreira19}.}

We can write all Wald statistics as%
\begin{equation*}
\widehat{\mathcal{W}}_{n}=\psi \left( R_{u},\widehat{D},\widehat{\Sigma }%
_{n},\widehat{\Phi }_{n}\right)
\end{equation*}%
(where we explicitly state the distribution of $R_{u}$ depends on the sample
size $n$). Its asymptotic behavior is given by%
\begin{equation*}
\mathcal{W}_{n}=\psi \left( \mathcal{R}_{u},\mathcal{D},\Sigma ,\Phi \right)
.
\end{equation*}%
where $\Phi =plim$ $\widehat{\Phi }_{n}=plim$ $n^{-1}V^{\prime }MV$ (if the
process is ergodic, $\Phi $ is just the variance of the reduced-form errors $%
V$). We then find the $1-\alpha $ quantile, say, $c_{\alpha }\left( d,\Sigma
,\Phi \right) $ of the null asymptotic distribution of%
\begin{equation*}
\psi \left( \mathcal{R}_{u},d,\Sigma ,\Phi \right) \text{, where }\mathcal{R}%
_{u}\sim N\left( 0,\Sigma _{uu}\right) .
\end{equation*}%
The final conditional test rejects the null when
\begin{equation*}
\widehat{\mathcal{W}}_{n}=\psi \left( R_{u},\widehat{D},\widehat{\Sigma }%
_{n},\widehat{\Phi }_{n}\right) >c_{\alpha }\left( \widehat{D},\widehat{%
\Sigma }_{n},\widehat{\Phi }_{n}\right) .
\end{equation*}

\section{Generalization to weakly-identified models (including GMM)} 

\label{sec:generalization}

Summarizing the setup in \citet{Andrews16}, it is assumed that there is a
sequence of models $F_{n}\left( \theta ,\gamma \right) $, which is indexed
by the sample size $n$. To illustrate the extension, we focus on a 
parameter of interest $\theta \in \Theta $, and presume there is an $l\times
1$ consistently estimable nuisance parameter $\gamma \in \Gamma $. The
objective is to test the null hypothesis $\theta =\theta _{0}$, presuming
the availability of three quantities: 1) a standardized sample moment vector
(or distance function) evaluated at the null, $h_{n}\left( \theta
_{0}\right) $\footnote{%
For the linear model, we can take either the (standardized) moment condition
$h_{n}\left( \beta \right) =\left( Z^{\prime }Z\right) ^{-1/2}Z^{\prime
}\left( y_{1}-Y_{2}\beta \right) $ or the distance function $h\left( \Pi
,\beta \right) =\left( Z^{\prime }Z\right) ^{-1}Z^{\prime }Y-\left[ \Pi
\beta :\Pi \right] $.}; 2) a sample gradient of $h_{n}\left( \theta \right) $
with respect to $\theta $ evaluated at the null, $\Delta h_{n}\left( \theta
_{0}\right) $; and 3) the consistent estimate $\hat{\gamma}$ for $\gamma $.

The main assumptions in \citet{Andrews16} are that for any true value $%
\left( \theta ,\gamma \right) \in \Theta \times \Gamma $:
\begin{equation*}
\left(
\begin{array}{c}
h_{n}\left( \theta _{0}\right)  \\
\Delta h_{n}\left( \theta _{0}\right)
\end{array}%
\right) \overset{d}{\rightarrow }\left(
\begin{array}{c}
h\left( \theta _{0}\right)  \\
\Delta h\left( \theta _{0}\right)
\end{array}%
\right)
\end{equation*}%
and
\begin{equation*}
\left(
\begin{array}{c}
h\left( \theta _{0}\right)  \\
vec\left( \Delta h\left( \theta _{0}\right) \right)
\end{array}%
\right) \sim N\left( \left(
\begin{array}{c}
m\left( \theta _{0}\right)  \\
vec\left( \mu \right)
\end{array}%
\right) ,\Sigma _{0}\right) \text{, where }\Sigma _{0}=\left(
\begin{array}{cc}
\Sigma _{hh} & \Sigma _{h\theta } \\
\Sigma _{\theta h} & \Sigma _{\theta \theta }%
\end{array}%
\right)
\end{equation*}%
and $\hat{\gamma}\overset{p}{\rightarrow }\gamma $. It is further assumed
that $\Sigma _{h\theta }$ and $\Sigma _{hh}$ are continuous in $\gamma $,
and hence consistently estimable.

The mean $m\left( \theta _{0}\right) $ belongs to a set $M\left( \mu ,\gamma
\right) \subseteq R^{k}$, with $\mu \in \mathcal{M}$, and is defined so that
when $\theta =\theta _{0}$, $m\left( \theta _{0}\right) =0$. The goal is to
test the null hypothesis $\left( m\left( \theta _{0}\right) ,\mu \right)
=\left( 0,\mu \right) $ against the alternative $\left( m\left( \theta
_{0}\right) ,\mu \right) =\left( \mathcal{M}\backslash \left\{ 0\right\}
,\mu \right) $ , for any unknown value of $\mu $.

We can once again consider the $k\times 1$ quantity
\begin{equation*}
\mathcal{D}=vec\left( \Delta h\left( \theta _{0}\right) \right) -\Sigma
_{\theta h}\Sigma _{hh}^{-1}h\left( \theta _{0}\right)
\end{equation*}%
which, by construction is independent of $h\left( \theta _{0}\right) $. The
Wald statistic based on the 2SLS estimator for the linear model simplifies to%
\begin{eqnarray*}
\widehat{\mathcal{W}}_{n} &=&R_{u}^{\prime }R_{2}\left[ R_{2}^{\prime
}\left( \widehat{b}_{n}\otimes I_{k}\right) ^{\prime }\widehat{\Sigma }%
_{0,n}\left( \widehat{b}_{n}\otimes I_{k}\right) R_{2}\right]
^{-1}R_{2}^{\prime }R_{u}\text{, where} \\
\widehat{b}_{n} &=&\left( 1,-R_{u}^{\prime }R_{2}\left( R_{2}^{\prime
}R_{2}\right) ^{-1}\right) ^{\prime }.
\end{eqnarray*}%
We can thus define a nonlinear analog as%
\begin{eqnarray*}
\widehat{\mathcal{W}}_{n} &=&h_{n}^{\prime }\Delta h_{n}\left[ \Delta
h_{n}^{\prime }\left( \widehat{b}_{n}\otimes I_{k}\right) ^{\prime }\widehat{%
\Sigma }_{n}\left( \widehat{b}_{n}\otimes I_{k}\right) \Delta h_{n}\right]
^{-1}\Delta h_{n}^{\prime }h_{n}\text{, where} \\
\widehat{b}_{n} &=&\left( 1,-h_{n}^{\prime }\Delta h_{n}\left( \Delta
h_{n}^{\prime }\Delta h_{n}\right) ^{-1}\right) ^{\prime }.
\end{eqnarray*}

(where we have suppressed the dependence on $\theta _{0}$), which will
converge in distribution to%
\begin{equation*}
\mathcal{W}\equiv \left( h^{\prime }\Delta h\right) \left[ \Delta h^{\prime
}\left( b\otimes I_{k}\right) ^{\prime }\Sigma \left( b\otimes I_{k}\right)
\Delta h\right] ^{-1}\left( \Delta h^{\prime }h\right)
\end{equation*}%
After substituting in $\Delta h=\mathcal{D}+\Sigma _{\theta h}\Sigma
_{hh}^{-1}h$, then it is easy to compute the $\left( 1-\alpha \right) $th
conditional quantile defined by
\begin{equation*}
\Pr \left[ \mathcal{W}>c\left( d,\Sigma ;\alpha \right) |\mathcal{D}=d\right]
=\alpha
\end{equation*}

The test is straightforward to implement as follows: reject the hypothesis
if and only if
\begin{equation*}
\widehat{\mathcal{W}}_{n}>c\left( \widehat{D}_{n},\widehat{\Sigma }%
_{n};\alpha \right) \text{, where }\widehat{D}_{n}=vec\left( \Delta
h_{n}\left( \theta _{0}\right) \right) -\widehat{\Sigma }_{\theta h}\widehat{%
\Sigma }_{hh}^{-1}h_{n}\left( \theta _{0}\right) .
\end{equation*}%
This test will have the property, under the null, that
\begin{equation*}
\lim \Pr \left[ \widehat{\mathcal{W}}_{n}>c\left( \widehat{D}_{n},\widehat{%
\Sigma }_{n};\alpha \right) |\widehat{D}_{n}=d\right] =\alpha
\end{equation*}%
for all values of $d$ and hence
\begin{equation*}
\lim \Pr \left[ \widehat{\mathcal{W}}_{n}>c\left( \widehat{D}_{n},\widehat{%
\Sigma }_{n};\alpha \right) \right] =\alpha
\end{equation*}%
as desired.

\section{Conclusion}\label{conclusion}

We are motivated by providing an inference method for researchers interested
in the over-identified linear instrumental variables model, and who have a
preference for using the 2SLS estimator $\hat{\beta}$ for inference, and who
do not wish to rely on the assumption of homoskedasticity. If errors are
assumed to be homoskedastic, one can use the results of %
\citet{StaigerStock97} and \citet{StockYogo05} to control the amount of
distortion in inference. As noted in \citet{AndrewsStockSun19}, the tables
in \citet{StockYogo05} do not apply to non-homoskedastic settings. %
\citet{Andrews18} provides a conservative two-step procedure that builds on %
\citet{StockYogo05} for more general DGPs.

To accommodate practitioners' preference for using the 2SLS estimator and
conventional robust standard errors, we present the robust Conditional Wald
(data-dependent) critical values for the Wald statistics robust to
heteroskedastic, autocorrelated, and/or clustered errors, which turns out to
be a relatively straightforward extension of the Conditional Wald test of %
\citet{Moreira03}; its derivation has been neglected in the weak-IV
literature, which has provided a number of other non-Wald procedures that
are both robust to non-homoskedastic errors and to arbitrarily weak
instruments.

Using existing results from the weak IV literature, we also generalize the
procedure to apply to the more general nonlinear models that are typically
estimated via minimum distance or GMM. We can explore several Wald
statistics within the nonlinear setup as well, contingent on the weights
employed in the criterion function. The final conditional test would
substitute the conventional critical value with a conditional quantile.

\newpage

{\normalsize
\begin{singlespace}

\bibliographystyle{aea}
\bibliography{References}

\end{singlespace}
}

\end{document}